\documentclass{ws-ijmpe}

\begin{document}

\markboth{Zhen Ouyang, Ju-Jun Xie, Hu-Shan Xu, Bing-Song
Zou}{Theoretical study on $pp \to pn \pi^+$ reaction at medium
energies}

\catchline{}{}{}{}{}

\title{THEORETICAL STUDY ON $pp \to pn \pi^+$ REACTION AT MEDIUM ENERGIES}

\author{\footnotesize ZHEN OUYANG}

\address{Institute of Modern Physics, CAS, Lanzhou 730000, China\\
Graduate University of Chinese Academy of Sciences, Beijing 100049, China\\
Theoretical Physics Center for Science Facilities, CAS, Beijing
100049, China\\ ouyangzh@impcas.ac.cn}

\author{JU-JUN XIE}

\address{Institute of High Energy Physics, CAS, Beijing 100049, China\\
Graduate University of Chinese Academy of Sciences, Beijing 100049, China\\
Theoretical Physics Center for Science Facilities, CAS, Beijing
100049, China\\
xiejujun@ihep.ac.cn}

\author{BING-SONG ZOU}

\address{Institute of High Energy Physics, CAS, Beijing 100049, China\\
Theoretical Physics Center for Science Facilities, CAS, Beijing
100049, China\\
zoubs@ihep.ac.cn}

\author{HU-SHAN XU}

\address{Institute of Modern Physics, CAS, Lanzhou 730000, China\\
Theoretical Physics Center for Science Facilities, CAS, Beijing
100049, China\\
hushan@impcas.ac.cn}

\maketitle

\begin{history}
\received{(received date)}
\revised{(revised date)}
\end{history}

\begin{abstract}
The $pp\to p n \pi^+$ reaction is a channel with the largest total
cross section for pp collision in COSY/CSR energy region.
 In this work, we investigate individual contributions
from various $N^*$ and $\Delta^{*}$ resonances with mass up to about
2 GeV for the $pp\to p n \pi^+$ reaction. We extend a resonance
model, which can reproduce the observed total cross section quite
well, to give theoretical predictions of various differential cross
sections for the present reaction at $T_p=2.88$ GeV. It could serve
as a reference for identifying new physics in the future experiments
at HIRFL-CSR.
\end{abstract}

\section{Introduction}
The study of excited nucleon states is very important for
understanding the internal structure of nucleon and the strong
interaction in the nonperturbative QCD domain~\cite{nisgur}. In the
early years, our investigation on the $N^{*}$ and $\Delta^{*}$
baryon spectroscopy was mainly based on $\pi N$ experiments, which
made observations unsatisfactory~\cite{liu,Capstick00}. An
outstanding problem is that, in many of its forms, the quark model
predicts a large amount of ``missing" $N^{*}$ and $\Delta^{*}$
states around 2 GeV/$c^2$, which have not to date been
observed~\cite{Capstick00,Isgur,Capstick92}. Therefore, it is of
necessity to search for these ``missing" $N^*$ and $\Delta^{*}$
states from other production processes. Moreover, even for those
well-established resonance states, properties like mass, width and
branching ratios still suffer large experimental
uncertainties~\cite{pdg2008}, which also need further studies in
more other production processes. Here we propose to look for
``missing" $N^*$ and $\Delta^{*}$ resonances in $pp\to pn \pi^+$
reaction. At COSY/J\"ulich, Experiments for studying $N^*$ and
$\Delta^{*}$ resonances through pp collisions are being carried out,
but there is a lack of a good $4\pi$ detector for complete
measurement of diverse differential cross sections. At present, a
heavy ion cooler-storage ring HIRFL-CSR---an accelerator system of
the same beam energy region with maximum incoming-proton kinetic
energy up to $2.88$ GeV~\cite{menu2004}, has already been completed
at Lanzhou. With its scheduled $4\pi$ hadron
detector~\cite{menu2004}, it will have a special advantage for
studying excited nucleon states through pp collisions.

Recently, BES collaboration has produced quite a few novel findings
on $N^*$ resonances by using various $N^*$ production processes from
$J/\psi$ or $\psi^\prime$
decays~\cite{besppe,besroper,yang,besprd74,besprd71}. Measurements
of the $J/\psi \to p \pi^- \bar n + c.c.$ decay by BES collaboration
showed two new, clear $N^*$ peaks in the $p\pi$ invariant mass
spectrum around 1360 MeV/$c^2$ and 2065 MeV/$c^2$,
respectively~\cite{besroper}. Of them the former one was identified
as the first direct observation of the $N^*(1440)$ peak in the $\pi
N$ invariant mass spectrum, which was confirmed by the CELSIUS-WASA
Collaboration in their $n\pi^+$ invariant mass spectrum of $pp\to pn
\pi^+$ reaction ~\cite{celsiusroper}. For the latter one, it is very
likely to be a long-sought missing $N^*$ peak around 2 GeV/$c^2$.
However, similar searches for it in $\psi'$ decays are
inconclusive~\cite{besprd74,besprd71}. Therefore, it is of necessity
to look for the new $N^*$ resonance in other reaction processes,
such as the $pp\to pn \pi^+$ reaction. Furthermore, in
Ref.~\cite{xie}, the authors found that the $\Delta^{*++}(1620)$
resonance gives an overwhelmingly large contribution in the $pp\to n
K^+ \Sigma^+$ reaction by t-channel $\rho^+$ exchange. If so, it is
also expected to make a significant contribution in the $pp\to pn
\pi^+$ reaction, as can be checked in the present work. In
Ref.~\cite{ouyang}, we have studied the $pp\to pn \pi^+$ reaction
for beam energies below 1.3 GeV. Here we extend the study of this
reaction to higher energies and investigate individual contributions
from various $N^*$ and $\Delta^{*}$ resonances with mass up to 2
GeV/$c^2$ for this reaction. We extend a resonance model, which can
describe the experimental data of the total cross section for beam
energies ranging from 0.8 GeV to 3.0 GeV quite well, to give
theoretical prediction of various differential cross sections  for
the $pp\to pn \pi^+$ reaction at $T_p=2.88$ GeV. It can be used for
the subsequent comparison with the experimental results at COSY and
HIRFL-CSR. Meanwhile, it could serve as a reference for the
construction of the scheduled $4\pi$ hadron detector at HIRFL-CSR.

\section{Formalism and ingredients}

\begin{figure}[htpb]
\begin{center}
\includegraphics[scale=0.5]{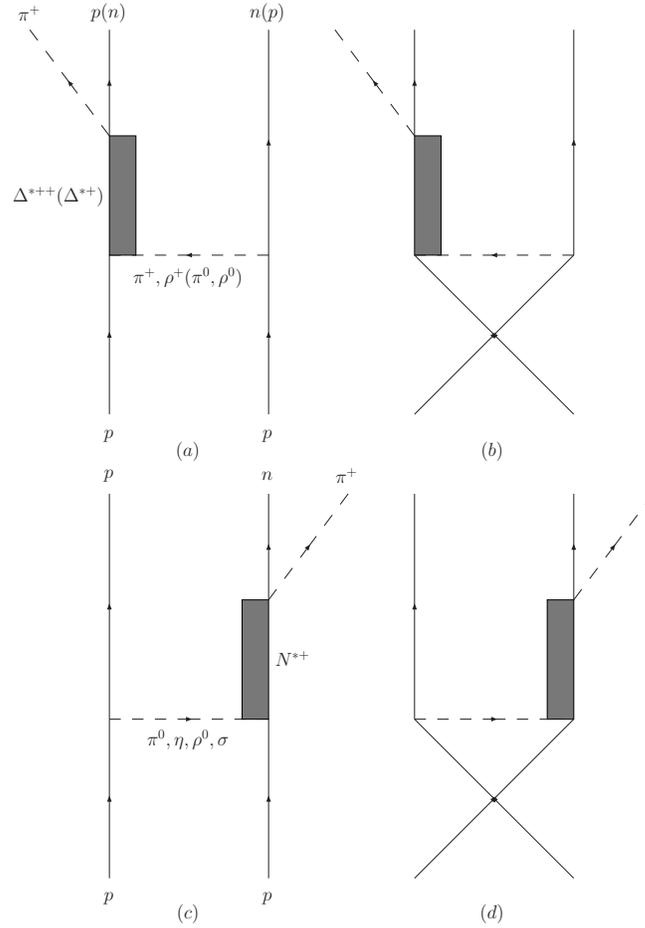} 
\caption{Feynman diagrams for $pp \to pn  \pi^+$ reaction. }
\label{diagram}
\end{center}
\end{figure}

We study the $pp \to pn\pi^+$ reaction within an effective
Lagrangian approach. In our model,all the mesons,baryons and
resonances are treated as fundamental fields. All the basic Feynman
diagrams involved in our calculation for this reaction are depicted
in Fig.~\ref{diagram}. In view of overall system invariant mass
about 3 GeV for $T_p=2.88$ GeV, we have checked contributions from
all the well-established $N^*$ and $\Delta^*$ resonances (overall
status 3 or 4 stars) below 2 GeV/$c^2$, but only present the results
of the relatively significant ones in next section. Meanwhile, we
investigate the contribution from $N^*(2065)$ resonance for the
present reaction. Explicitly, we list in Table~\ref{table} all the
$N^*$, $\Delta^*$ resonances and the meson exchanges considered in
our present calculation.

For $N^*(2065)$, according to results in Ref.~\cite{besroper}, its
spin-parity is limited to be ${1/2}^+$ and ${3/2}^+$, and it is more
likely that both are needed. In the quark model there are
predictions for the existence of $N^*$ resonances with spin-parity
${1/2}^+$ and ${3/2}^+$ between 2.0 and 2.1
GeV/$c^2$~\cite{Capstick00,Isgur,Capstick92}. Since the spin-parity
of the new resonance(s) was not well determined, we assume that this
peak consists of exactly those resonances with $J^p={1/2}^+,{3/2}^+$
predicted in Ref.~\cite{Capstick92}, which are
$N^*(1975)$($J^p={1/2}^+$), $N^*(2030)$($J^p={3/2}^+$) and
$N^*(2065)$($J^p={1/2}^+$). Among them $N^*(2065)$ has much stronger
coupling to $\pi N$ than the other two, which is in accord with BES
results. As in Ref.~\cite{tsushima96}, here we also treat the
observed $N^*(2065)$ peak as an effective $N^*(2065)$($J^p={1/2}^+$)
resonance which represents all contributions of the three resonances
$N^*(1975)$($J^p={1/2}^+$), $N^*(2030)$($J^p={3/2}^+$) and
$N^*(2065)$($J^p={1/2}^+$). In so doing, the coupling constant
${g^2}_{\pi N N^*(2065)}/{4\pi}$ is scaled by a factor of 1.122. See
Ref.~\cite{tsushima96} for details of this effective treatment. Of
course, we have used the BES observed values for the mass and width
of $N^*(2065)$ to determine its relevant coupling constant, see
Table~\ref{table}. In view of scanty information for its decay
branching fractions, here we regard $N \pi$ as the dominant decay
mode of $N^*(2065)$, whereas the $N^*(2065)$ peak in invariant mass
$M_{p\pi^-}$ spectrum is so strong and highly significant. So, in
our calculations we have taken an artificial branching ratio up to 1
for $N \pi$ decay mode.

The effective Lagrangian densities involved for describing the
meson-$NN$ vertices are:
\begin{equation}
{\cal L}_{\pi N N}  = - \frac{f_{\pi N N}}{m_{\pi}} \overline{u}_{N}
\gamma_5 \gamma_{\mu} \vec\tau \cdot \partial^{\mu} \vec\psi_{\pi}
u_N , \label{piNN}
\end{equation}
\begin{equation}
{\cal L}_{\eta N N}  = -i g_{\eta N N} \overline{u}_{N} \gamma_5
 \psi_{\eta} u_N ,
\label{etaNN}
\end{equation}
\begin{equation}
{\cal L}_{\sigma N N}  =  g_{\sigma N N}  \overline{u}_{N}
\psi_{\sigma} u_N, \label{sigNN}
\end{equation}
\begin{equation}
{\cal L}_{\rho N N}  = - g_{\rho N N} \overline{u}_{N}
(\gamma_{\mu}+\frac{\kappa}{2m_N} \sigma_{\mu \nu}
\partial^{\nu}) \vec\tau \cdot  \vec\psi_{\rho}^{\mu} u_N.
\label{rhoNN}
\end{equation}

At each vertex a relevant off-shell form factor is used. In our
computation, we take the same form factors as used in the well-known
Bonn potential model~\cite{mach}:
\begin{equation}
F^{NN}_M(k^2_M)=(\frac{\Lambda^2_M-m_M^2}{\Lambda^2_M- k_M^2})^n
\end{equation}
with n=1 for $\pi$, $\eta$ and $\sigma$ mesons and n=2 for $\rho$
meson. $k_M$, $m_M$ and $\Lambda_M$ are the 4-momenta, mass and
cut-off parameter for the exchanged meson ($M$), respectively. The
coupling constants and the cut-off parameters are taken as the
following ones~\cite{xie,ouyang,mach,tsushima,sibi}: $g_{\pi
NN}^2/4\pi=14.4$, $g_{\eta NN}^2/4\pi=0.4$, $\Lambda_{\pi}=
\Lambda_{\eta}=1.3$ GeV, $g_{\sigma NN}^2/4\pi=5.69$,
$\Lambda_{\sigma}=2.0$ GeV, $g_{\rho NN}^2/4\pi=0.9$,
$\Lambda_{\rho}=1.85$ GeV, and $\kappa=6.1$. Note that the constant
$g_{\pi NN}$ is related to $f_{\pi NN}$ of Eq.(\ref{piNN}) as
$g_{\pi NN}=(f_{\pi N N}/m_{\pi})2m_N$~\cite{scadron}.

To calculate the amplitudes of diagrams in Fig.~\ref{diagram} within
the resonance model, we also need to know interaction vertices
involving $N^*$ and $\Delta^*$ resonances. In Ref.~\cite{zouprc03},
a Lorentz covariant orbital-spin scheme for $N^* NM$ couplings has
been described in detail, which can be easily extended to describe
all the couplings that appear in the Feynman diagrams depicted in
Fig.~\ref{diagram}. By using that scheme, we can easily obtain the
effective couplings:
\begin{eqnarray}
{\cal L}_{\pi N \Delta(1232)} &=& g_{\Delta(1232) N \pi}
\overline{u}_{N}  \partial^{\mu} \psi_{\pi} u_{\Delta(1232)\mu} + \text{h.c.}, \label{1232pi} \\
{\cal L}_{\pi N N^*(1440)} &=& g_{N^*(1440)N \pi}
\overline{u}_{N} \gamma_5 \gamma_{\mu}  \partial^{\mu} \psi_{\pi} u_{N^*(1440)} + \text{h.c.}, \label{1440pi} \\
{\cal L}_{\sigma N N^*(1440)} &=& g_{N^*(1440) N \sigma}
\overline{u}_{N} \psi_{\sigma} u_{N^*(1440)} + \text{h.c.},
\label{1440sig} \\
{\cal L}_{\pi N N^*(1520)} &=& g_{N^*(1520)N \pi} \overline{u}_{N}
\gamma_5 \gamma_{\mu}  p_{\pi}^{\mu}p_{\pi}^{\nu} \psi_{\pi}
u_{N^*(1520)\nu} + \text{h.c.},
\label{1520pi} \\
{\cal L}_{\rho N N^*(1520)} &=& g_{N^*(1520)N \rho}
\overline{u}_{N}  \psi_{\rho}^{\mu} u_{N^*(1520)\mu} + \text{h.c.}, \label{1520rho} \\
{\cal L}_{\pi N N^*(1535)} &=& g_{N^*(1535) N \pi} \overline{u}_{N}
\psi_{\pi} u_{N^*(1535)} + \text{h.c.},
\label{1535pi} \\
{\cal L}_{\eta N N^*(1535)} &=& g_{N^*(1535) N \eta}
\overline{u}_{N} \psi_{\eta} u_{N^*(1535)} + \text{h.c.},
\label{1535eta} \\
{\cal L}_{\rho N N^*(1535)} &=& g_{N^*(1535) N \rho}
\overline{u}_{N} \gamma_5
(\gamma_{\mu}-\frac{q_{\mu}\gamma^{\nu}q_{\nu}}{q^2})
\psi_{\rho}^{\mu} u_{N^*(1535)}
 + \text{h.c.}, \label{1535rho}\\
{\cal L}_{\pi N \Delta^*(1600)} &=& g_{\Delta^*(1600) N \pi}
\overline{u}_{N}  \partial^{\mu} \psi_{\pi} u_{\Delta^*(1600)\mu} + \text{h.c.}, \label{1600pi} \\
{\cal L}_{\pi N \Delta^*(1620)} &=& g_{\Delta^*(1620) N \pi}
\overline{u}_{N}   \psi_{\pi} u_{\Delta^*(1620)} + \text{h.c.}, \label{1620pi} \\
{\cal L}_{\rho N \Delta^*(1620)} &=& g_{\Delta^*(1620) N \rho}
\overline{u}_{N} \gamma_5 (\gamma_{\mu}-\frac{q_{\mu}\gamma^{\nu}q_{\nu}}{q^2}) \psi_{\rho}^{\mu} u_{\Delta^*(1620)} + \text{h.c.}, \label{1620rho} \\
{\cal L}_{\pi N N^*(1650)} &=& g_{N^*(1650) N \pi} \overline{u}_{N}
\psi_{\pi} u_{N^*(1650)} + \text{h.c.},
\label{1650pi} \\
{\cal L}_{\eta N N^*(1650)} &=& g_{N^*(1650) N \eta}
\overline{u}_{N} \psi_{\eta} u_{N^*(1650)} + \text{h.c.},
\label{1650eta} \\
{\cal L}_{\rho N N^*(1650)} &=& g_{N^*(1650) N \rho}
\overline{u}_{N} \gamma_5
(\gamma_{\mu}-\frac{q_{\mu}\gamma^{\nu}q_{\nu}}{q^2})
\psi_{\rho}^{\mu} u_{N^*(1650)}
 + \text{h.c.}, \label{1650rho} \\
{\cal L}_{\pi N N^*(1675)} &=& g_{N^*(1675)N \pi}
\overline{u}_{N}   p_{\pi}^{\mu}p_{\pi}^{\nu} \psi_{\pi} u_{N^*(1675)\mu\nu} + \text{h.c.}, \label{1675pi} \\
{\cal L}_{\pi N N^*(1680)} &=& g_{N^*(1680)N \pi} \overline{u}_{N}
\gamma_5 \gamma_{\mu}  p_{\pi}^{\mu}p_{\pi}^{\nu}p_{\pi}^{\lambda}
\psi_{\pi} u_{N^*(1680)\nu\lambda}
+ \text{h.c.}, \label{1680pi} \\
{\cal L}_{\pi N N^*(1700)} &=& g_{N^*(1700)N \pi} \overline{u}_{N}
\gamma_5 \gamma_{\mu}  p_{\pi}^{\mu}p_{\pi}^{\nu} \psi_{\pi}
u_{N^*(1700)\nu}
+ \text{h.c.}, \label{1700pi} \\
{\cal L}_{\rho N N^*(1700)} &=& g_{N^*(1700)N \rho}
\overline{u}_{N}  \psi_{\rho}^{\mu} u_{N^*(1700)\mu} + \text{h.c.}, \label{1700rho} \\
{\cal L}_{\pi N \Delta^*(1700)} &=& g_{\Delta^*(1700)N \pi}
\overline{u}_{N} \gamma_5 \gamma_{\mu}  p_{\pi}^{\mu}p_{\pi}^{\nu}
\psi_{\pi} u_{\Delta^*(1700)\nu}
 + \text{h.c.}, \label{delta1700pi} \\
{\cal L}_{\rho N \Delta^*(1700)} &=& g_{\Delta^*(1700)N \rho}
\overline{u}_{N}  \psi_{\rho}^{\mu} u_{\Delta^*(1700)\mu} + \text{h.c.}, \label{delta1700rho} \\
{\cal L}_{\pi N N^*(1710)} &=& g_{N^*(1710)N \pi}
\overline{u}_{N} \gamma_5 \gamma_{\mu}  \partial^{\mu} \psi_{\pi} u_{N^*(1710)} + \text{h.c.}, \label{1710pi} \\
{\cal L}_{\eta N N^*(1710)} &=& g_{N^*(1710)N \eta}
\overline{u}_{N} \gamma_5 \gamma_{\mu}  \partial^{\mu} \psi_{\eta} u_{N^*(1710)} + \text{h.c.}, \label{1710eta} \\
{\cal L}_{\sigma N N^*(1710)} &=& g_{N^*(1710) N \sigma}
\overline{u}_{N} \psi_{\sigma} u_{N^*(1710)} + \text{h.c.},
\label{1710sig} \\
{\cal L}_{\pi N N^*(1720)} &=& g_{N^*(1720) N \pi}
\overline{u}_{N}  \partial^{\mu} \psi_{\pi} u_{N^*(1720)\mu} + \text{h.c.}, \label{1720pi} \\
{\cal L}_{\eta N N^*(1720)} &=& g_{N^*(1720) N \eta}
\overline{u}_{N}  \partial^{\mu} \psi_{\eta} u_{N^*(1720)\mu} +
\text{h.c.}, \label{1720eta}\\
{\cal L}_{\pi N \Delta^*(1905)} &=& g_{\Delta^*(1905)N \pi}
\overline{u}_{N} \gamma_5 \gamma_{\mu}
p_{\pi}^{\mu}p_{\pi}^{\nu}p_{\pi}^{\lambda}
\psi_{\pi} u_{\Delta^*(1905)\nu\lambda} + \text{h.c.}, \label{1905pi} \\
{\cal L}_{\pi N \Delta^*(1910)} &=& g_{\Delta^*(1910)N \pi}
\overline{u}_{N} \gamma_5 \gamma_{\mu}  \partial^{\mu} \psi_{\pi}
u_{\Delta^*(1910)} + \text{h.c.}, \label{1910pi}\\
{\cal L}_{\rho N\Delta^*(1910)} &=& g_{\Delta^*(1910) N \rho}
\overline{u}_{N} (p_{N\mu}\!-\!k_{\rho
\mu}\!-\!\frac{(m_N^2-k_{\rho}^2)q_{\mu}}{q^2}) \psi_{\rho}^{\mu}
u_{\Delta^*(1910)} + \text{h.c.}, \label{1910rho} \\
{\cal L}_{\pi N\Delta^*(1920)} &=& g_{\Delta^*(1920) N \pi}
\overline{u}_{N}  \partial^{\mu} \psi_{\pi} u_{\Delta^*(1920)\mu} + \text{h.c.}, \label{1920pi} \\
{\cal L}_{\pi N \Delta^*(1930)} &=& g_{\Delta^*(1930)N \pi}
\overline{u}_{N}   p_{\pi}^{\mu}p_{\pi}^{\nu} \psi_{\pi} u_{\Delta^*(1930)\mu\nu} + \text{h.c.}, \label{1930pi} \\
{\cal L}_{\pi N N^*(2065)} &=& g_{N^*(2065)N \pi} \overline{u}_{N}
\gamma_5 \gamma_{\mu}  \partial^{\mu} \psi_{\pi} u_{N^*(2065)} +
\text{h.c.}. \label{2065pi}
\end{eqnarray}

For the relevant vertices involving $N^*$ and $\Delta^*$ resonances,
the off-shell form factors are adopted as follows:
\begin{equation}
F_M(k^2_M)=(\frac{\Lambda^{*2}_M-m_M^2}{\Lambda^{*2}_M- k_M^2})^n
\end{equation}
where n=1 for all the resonances except for n=2 for $\Delta(1232)$.
All the coupling constants and cut-off parameters used in the
present paper are listed in Table~\ref{table}. In addition, we also
introduce form factors for the off-shell baryon resonances as in
Refs.~\cite{pennerprc66,shkprc72,feusterprc58}
\begin{equation}
F_R(q)=\frac{\Lambda^4}{\Lambda^4+ (q^2-m_R^2)^2},
\end{equation}
with $\Lambda$= 0.8 GeV.

\begin{table}[htbp]
\caption{Relevant parameters of $N^*$ and $\Delta^*$ included in our
calculations. The widths and branching ratios are taken from PDG$^6$
and the cut-off parameters are from Refs.$^{14,15,18,25}$. Here the
$g^2/4\pi$ for $N^*(2065) N\pi$ vertex has already been scaled by a
factor of 1.122.} \label{table}
\begin{tabular}{ccccccc}
\hline\hline
Resonance   &Width/GeV & Decay mode & Branching ratio &$g^2/4\pi$ & Cut-off/GeV\\
 \hline
$\Delta(1232)$ & 0.118& $N\pi$& 1.0& 19.54& 0.6 \\
$N^*(1440)$  & 0.3& $N\pi$& 0.65& 0.51& 1.3\\
     &  & $N\sigma$& 0.075&3.20& 1.1\\
$N^*(1520)$  & 0.115& $N\pi$& 0.6& 1.73& 0.8\\
     &  & $N\rho$& 0.09&1.32& 0.8\\
$N^*(1535)$  & 0.15& $N\pi$& 0.45& 0.037& 1.3\\
     &  & $N\eta$& 0.525&0.34& 1.3\\
     &  & $N\rho$& 0.02& 0.097& 1.3\\
$\Delta^*(1600)$  & 0.35& $N\pi$& 0.175& 1.09& 0.8\\
$\Delta^*(1620)$  & 0.145& $N\pi$& 0.25& 0.06& 1.3\\
     &  & $N\rho$& 0.14&0.37& 1.3\\
$N^*(1650)$  & 0.165 & $N\pi$ &0.775 & 0.06& 0.8\\
     &  & $N\eta$& 0.065&0.026& 0.8\\
     &  & $N\rho$& 0.02&0.015& 0.8\\
$N^*(1675)$  & 0.15 & $N\pi$ &0.4 & 2.16& 0.8\\
$N^*(1680)$  & 0.13 & $N\pi$ &0.675 & 5.53& 0.8\\
$N^*(1700)$  & 0.1& $N\pi$& 0.1& 0.075& 0.8\\
     &  & $N\rho$& 0.07& 0.043& 0.8\\
$\Delta^*(1700)$  & 0.3& $N\pi$& 0.15& 1.02& 0.8\\
     &  & $N\rho$& 0.125& 0.69& 0.8\\
$N^*(1710)$  & 0.1 & $N\pi$ &0.15 & 0.012& 0.8\\
     &  & $N\eta$& 0.062&0.042& 0.8\\
     &  & $N\sigma$& 0.25&0.085& 1.1\\
$N^*(1720)$  & 0.2 & $N\pi$ &0.15 & 0.12& 0.8\\
     &  & $N\eta$& 0.04&0.28& 0.8\\
$\Delta^*(1905)$  & 0.33& $N\pi$& 0.12& 1.74& 0.8\\
$\Delta^*(1910)$  & 0.25& $N\pi$& 0.225& 0.076& 0.8\\
     &  & $N\rho$& 0.37&0.29& 0.8\\
$\Delta^*(1920)$  & 0.2& $N\pi$& 0.125& 0.18& 0.8\\
$\Delta^*(1930)$  & 0.36& $N\pi$& 0.1& 1.06& 0.8\\
$N^*(2065)$  & 0.165 & $N\pi$ &$\sim$1.0 & 0.057& 1.3\\
\hline\hline
\end{tabular}
\end{table}

The propagators can be written as
\begin{equation}
G(q) =  \frac{\not\! q + m_R}{q^2 - m^2_R + im_R \Gamma_R}\,
\label{spin1/2}
\end{equation}
for the spin-$\frac{1}{2}$ resonances,
\begin{equation}
G_{\mu \nu}(q) =  \frac{-P_{\mu \nu}(q)}{q^2 - m^2_R + im_R
\Gamma_R}\,  \label{spin3/2}
\end{equation}
with
\begin{eqnarray}
P_{\mu \nu}(q)  = -  (\not\! q + m_R)  [ g_{\mu \nu} - \frac{1}{3}
\gamma_\mu \gamma_\nu - \frac{1}{3 m_R}( \gamma_\mu q_\nu -
\gamma_\nu q_\mu) - \frac{2}{3 m^2_R} q_\mu
q_\nu] ,\nonumber\\
&& \label{pmunu}
\end{eqnarray}
for the spin-$\frac{3}{2}$ resonances, and
\begin{equation}
G_{\mu \nu \alpha \beta}(q) =  \frac{-P_{\mu \nu \alpha
\beta}(q)}{q^2 - m^2_R + im_R \Gamma_R}\,  \label{spin-3/2}
\end{equation}
with
\begin{eqnarray}
P_{\mu \nu \alpha \beta}(q)  = &-&  (\not\! q + m_R)  [
\frac{1}{2}(\tilde g_{\mu\alpha}\tilde g_{\nu\beta}+\tilde
g_{\mu\beta}\tilde g_{\nu\alpha} )-\frac{1}{5}\tilde
g_{\mu\nu}\tilde
g_{\alpha\beta}\\
&+&\frac{1}{10}(\tilde\gamma_\mu\tilde\gamma_\alpha\tilde
g_{\nu\beta}+\tilde\gamma_\nu\tilde\gamma_\beta\tilde
g_{\mu\alpha}+\tilde\gamma_\mu\tilde\gamma_\beta\tilde
g_{\nu\alpha}+\tilde\gamma_\nu\tilde\gamma_\alpha\tilde g_{\mu\beta}
)] , \label{pmunualphabeta}
\end{eqnarray}
\begin{equation}
\tilde g_{\mu\nu}(q) = -g_{\mu\nu}+{q_\mu q_\nu\over m^2_R}, \quad
\tilde\gamma_\mu = -\gamma_\mu + {\not\! q q_\mu\over m^2_R}.
\end{equation}
for the spin-$\frac{5}{2}$ resonances.

After the effective Lagrangians, coupling constants and propagators
fixed, the amplitudes for various diagrams can be written down
straightforwardly by following the Feynman rules. And the total
amplitude is just their simple sum. Here we give explicitly the
individual amplitude corresponding to $N^*(1440)(\pi^0$ exchange),
as an example,
\begin{eqnarray}
{\cal M}(N^{*}(1440),\pi^0)& = &\sqrt{2} \frac{f_{\pi N N}}{m_{\pi}}
g^2_{N^* N \pi} \bar{u}_n(p_n,s_n) \gamma_5 \not \! p_{\pi}
G^{N^*(1440)}(q) \gamma_5 \not\! k_{\pi}
u_1(p_1,s_1) \frac{i}{k^2_{\pi}-m^2_{\pi}}   \nonumber\\
&& \bar{u}_3(p_3,s_3)  \gamma_5 \not\! k_{\pi} u_2(p_2,s_2) + (\text
{exchange term with } p_1 \leftrightarrow p_2),
\end{eqnarray}
where $u_n(p_n,s_n)$, $u_3(p_3,s_3)$, $u_1(p_1,s_1)$, $u_2(p_2,s_2)$
denote the spin wave functions of the outgoing neutron, proton in
the final state and two initial protons, respectively. $p_{\pi}$ and
$k_{\pi}$ are the 4-momenta of the outgoing and the exchanged pion
mesons. $p_{1}$ and $p_{2}$ represent the 4-momenta of the two
initial protons. The coupling constant appearing herein can be
determined from the experimentally observed partial decay width of
$N^*(1440)$ resonance as follows,
\begin{equation}
\Gamma_{N^*(1440)\to N \pi}=\frac{3g^2_{N^*(1440) N
\pi}p_N^{cm}}{4\pi
}[\frac{m^2_\pi(E_N-m_N)}{m_{N^*(1440)}}+2(p_N^{cm})^2],
\label{1440d}
\end{equation}
with
\begin{equation}
p_N^{cm}=\sqrt{\frac{(m^2_{N^*(1440)}-(m_N+m_{\pi})^2)
(m^2_{N^*(1440)}-(m_N-m_{\pi})^2)}{4m^2_{N^*(1440)}}},
\end{equation}
\begin{equation}
E_N=\sqrt{(p_N^{cm})^2+m^2_N}.
\end{equation}
All the other coupling constants can be obtained similarly. See
Refs.~\cite{xie,ouyang} for details.

Then the calculation of the cross section $\sigma (pp \to pn \pi^+)$
is straightforward,
\begin{eqnarray}
d\sigma (pp\to p n \pi^+)=\frac{1}{4}\frac{m^2_p}{F} \sum_{s_i,s_f}
|{\cal M}|^2\frac{m_p d^{3} p_{3}}{E_{3}} \frac{d^{3} p_{\pi}}{2
E_{\pi}} \frac{m_{n} d^{3} p_{n}}{E_{n}} \delta^4
(p_1\!+\!p_2\!-\!p_3\!-\!p_{\pi}\!-\!p_{n})\nonumber\\  \label{eqcs}
\end{eqnarray}
with the flux factor
\begin{eqnarray}
F=(2 \pi)^5\sqrt{(p_1\cdot p_2)^2-m^4_p}. \label{eqff}
\end{eqnarray}
The factors 1/4 and $\sum_{s_i,s_f}$ emerge for the simple reason
that the polarization of initial and final particles is not
considered.

\section{Numerical results and discussion}

\noindent

\begin{figure}[htbp]
\begin{center}
\includegraphics[scale=0.59]{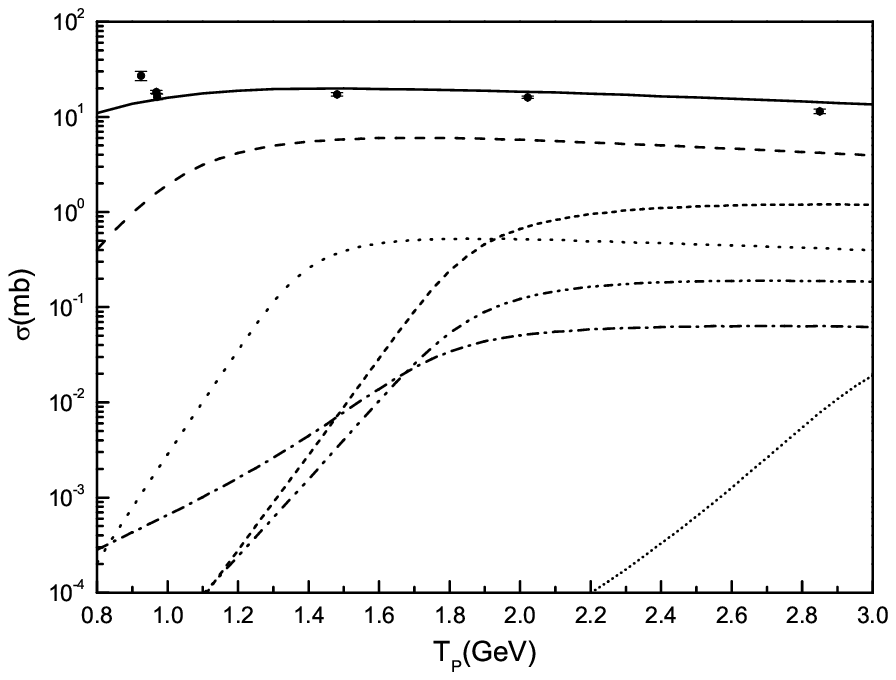}
\includegraphics[scale=0.59]{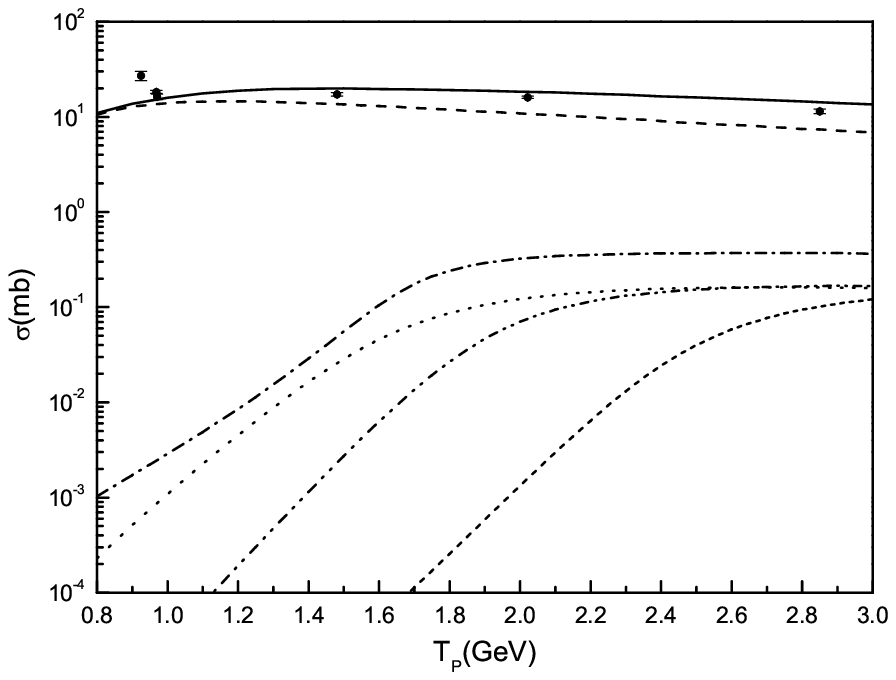}
\caption{Total cross section and contributions from various $N^*$
(left) and $\Delta^*$ (right) resonances as a function of T$_\text
P$ for the $pp \to pn \pi^+$ reaction with the solid line as the
simple incoherent sum of all contributions, compared with
data$^{26}$. Left: the dashed, dotted, dot-dashed, dot-dot-dashed,
short-dashed and short-dotted lines represent individual
contributions from $N^*(1440)$, $N^*(1520)$, $N^*(1650)$,
$N^*(1675)$, $N^*(1680)$, and $N^*(2065)$, respectively. Right: the
dashed, dotted, dot-dashed, dot-dot-dashed, and short-dashed lines
represent individual contributions from $\Delta(1232)$,
$\Delta^*(1600)$, $\Delta^*(1620)$, $\Delta^*(1700)$, and
$\Delta^*(1905)$,
respectively.} %
\label{fig1}%
\end{center}
\end{figure}

With the formalism and ingredients discussed in the former section,
we computed the total cross section versus the kinetic energy of the
proton beam (T$_\text P$) for the $pp \to pn  \pi^+$ reaction by
using a Monte Carlo multi-particle phase space integration program.
The results for T$_\text P$ ranging from 0.8 to 3.0 GeV are shown in
Fig.~\ref{fig1} along with experimental data~\cite{data} for
comparison.

\noindent

\begin{figure}[htbp]
\begin{center}
\includegraphics[scale=0.6]%
{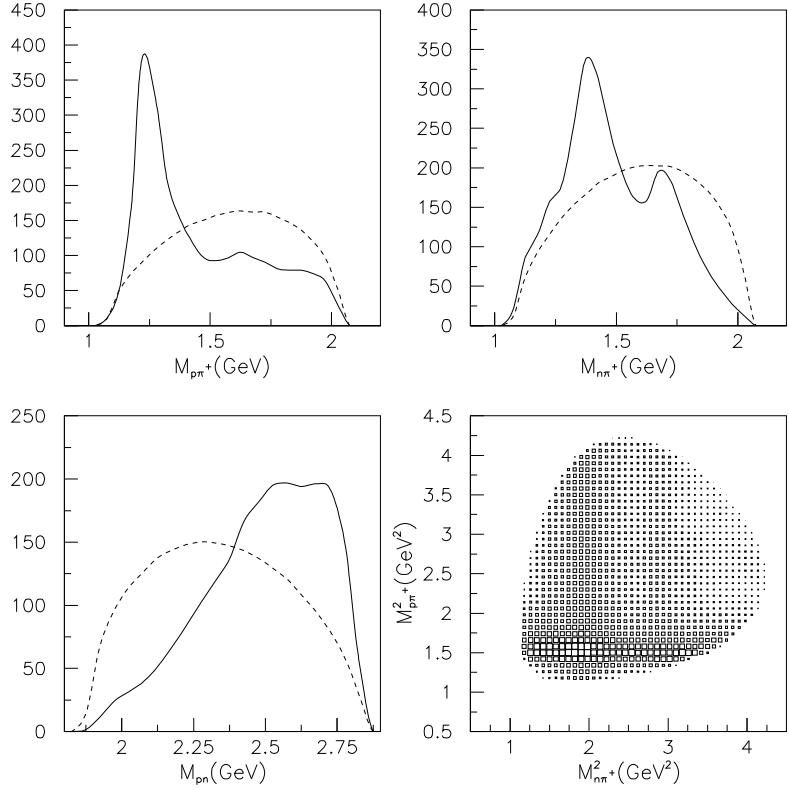}%
\caption{Theoretical prediction of invariant mass spectra and Dalitz
plot for $pp \to pn \pi^+$ reaction at T$_\text P=2.88$ GeV. The
dashed curves stand for pure
phase space distribution while the solid curves include the interaction amplitudes.}%
\label{pnpi}%
\end{center}
\end{figure}

\begin{figure}[htbp]
\begin{center}
\includegraphics[scale=0.6]%
{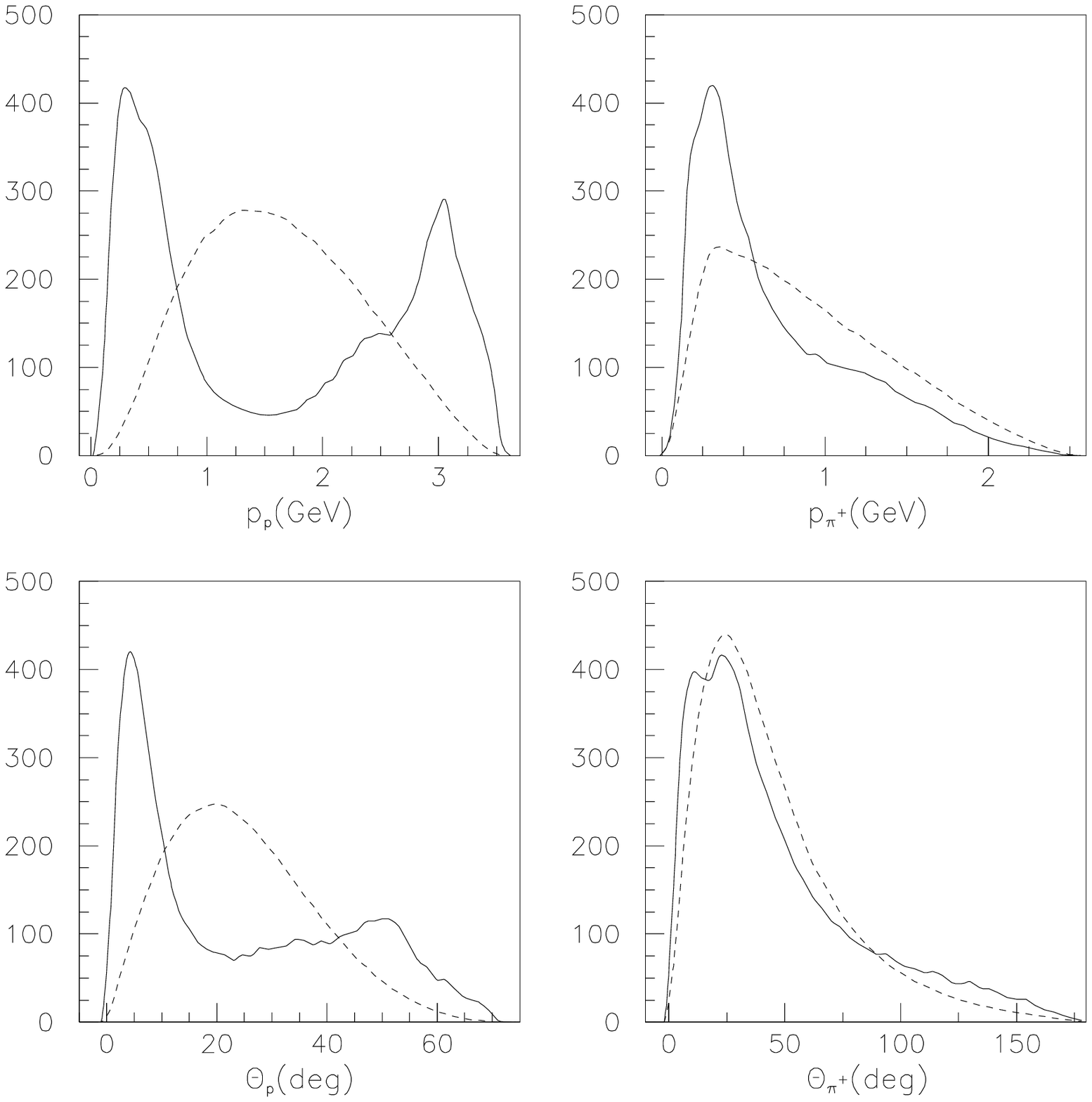}%
\caption{The momentum and angular distributions of the final proton
and charged pion for $pp \to pn \pi^+$ reaction at T$_\text P=2.88$
GeV, compared with pure
phase space distributions (dashed curves).}%
\label{pnpi2}%
\end{center}
\end{figure}

As one can see from Fig.~\ref{fig1}, the experimental data of total
cross section are reproduced reasonably well by our theoretical
calculations over the entire energy range. Note that we have
considered the interference terms between the direct amplitudes
(diagram a,c in Fig.~\ref{diagram}) and the corresponding exchange
amplitudes (diagram b,d in Fig.~\ref{diagram}) in our calculations.
However,the interference terms between different
resonance-excitation processes and between various meson-exchange
diagrams are ignored. We also show contributions of various
components which are large and not negligible there, $N^*$
contributions in Fig.~\ref{fig1} (left) and $\Delta^*$ contributions
in Fig.~\ref{fig1} (right), respectively. Individual contributions
from $N^*(1440)$, $N^*(1520)$, $N^*(1650)$, $N^*(1675)$,
$N^*(1680)$, and $N^*(2065)$ are presented in Fig.~\ref{fig1} (left)
by dashed, dotted, dot-dashed, dot-dot-dashed, short-dashed and
short-dotted lines, respectively. And contributions from
$\Delta(1232)$, $\Delta^*(1600)$, $\Delta^*(1620)$,
$\Delta^*(1700)$, and $\Delta^*(1905)$ are shown in Fig.~\ref{fig1}
(right) by dashed, dotted, dot-dashed, dot-dot-dashed, and
short-dashed lines, respectively. One can find that contributions
from $\Delta(1232)$ and $ N^*(1440)$ are still dominant in present
energy region and the contribution from $ N^*(1680)$ becomes
significant for kinetic energy above 2.0 GeV. We also give our
predictions of invariant mass spectra and Dalitz plot in
Fig.~\ref{pnpi} and the momentum and angular distributions of the
final charged particles in Fig.~\ref{pnpi2} for $pp \to pn  \pi^+$
reaction at T$_\text P=2.88$ GeV.

The $pp\to p n \pi^+$ reaction is a channel with the largest total
cross section for pp collision in the present energy region. Since
the kinetic energy T$_\text P$ of the proton beam at HIRFL-CSR can
reach $2.88$ GeV with luminosity above
$10^{32}cm^{-2}s^{-1}$~\cite{menu2004}, the event rate will be so
large that it is easy to collect enough events and to get high
statistics data for this channel at HIRFL-CSR. The theoretical
prediction of various observables given in this work would provide
us more knowledge on the relevant physics when compared to the
future experimental results. Meanwhile, a good measurement of
invariant mass spectra and Dalitz plot will play the key role for
identifying new resonances and determining their parameters. In this
regard, the scheduled $4\pi$ hadron detector at HIRFL-CSR will be
particularly competent.

As mentioned above, the spin-parity of the $N^*(2065)$ peak was not
well determined by the BES collaboration, therefore we have used an
effective treatment for the $N^*(2065)$ resonance(s) as the authors
did in Ref.~\cite{tsushima96}. This effective description would
indeed be a very good approximation for the total cross section, but
generally speaking, it might reproduce the differential cross
sections not so well. However, due to the dominance of $N^*(2065)$
(its magnitude is much stronger than the other two) among the three
resonances predicted in Ref.~\cite{Capstick92}, hence even for the
description of various differential cross sections, it would be
acceptable. Of course, this issue still waits for an exact answer
from future experimental results at HIRFL-CSR.

To sum up, in this paper we investigate individual contributions
from diverse $N^*$ and $\Delta^{*}$ resonances up to 2 GeV/$c^2$ for
the $pp\to p n \pi^+$ reaction. We extend a resonance model, which
can describe the observed total cross section for beam energies
ranging from 0.8 GeV to 3.0 GeV quite well, to give theoretical
prediction of various differential cross sections  for this reaction
at $T_p=2.88$ GeV. It can be used for identifying new physics in the
future experiments at HIRFL-CSR. It could also serve as a reference
for the construction of the scheduled $4\pi$ hadron detector at
HIRFL-CSR, which is quite possible to offer more physical
information and to help us understanding the relevant physics
better.

\bigskip
\noindent

{\bf Acknowledgements:} We thank C. Zheng and B.C. Liu for useful
discussions. This work is partly supported by the National Natural
Science Foundation of China under grants Nos. 10435080, 10521003,
10635080, and by the Chinese Academy of Sciences under project No.
KJCX2-SW-N18, KJCX3-SYW-N2,CXTD-J2005-1.

\end{document}